\documentclass[preprint,12pt]{elsarticle}

\usepackage{amssymb}
\usepackage{amsmath}
\usepackage{booktabs}
\usepackage{amsthm}

\makeatletter
\def\ps@pprintTitle{%
   \let\@oddhead\@empty
   \let\@evenhead\@empty
   \def\@oddfoot{\reset@font\hfil\thepage\hfil}
   \let\@evenfoot\@oddfoot
}
\makeatother

\begin{document}

\begin{frontmatter}



\title{A physics-informed reinforcement learning approach for the interfacial area transport in two-phase flow}

\author[mymainaddress]{Zhuoran Dang\corref{correspondingauthor}}
\ead{zdang@purdue.edu} 
\author[mymainaddress]{Mamoru Ishii} 
\cortext[correspondingauthor]{Corresponding author}
\address[mymainaddress]{School of Nuclear Engineering, Purdue University, 516 Northwestern Ave., West Lafayette, IN, 47907, USA}

\begin{abstract}
The prediction of interfacial structure in two-phase flow systems is difficult and challenging. In this paper, a novel physics-informed reinforcement learning-aided framework (PIRLF) for the interfacial area transport is proposed. A Markov Decision Process that describes the bubble transport is established by assuming that the development of two-phase flow is a stochastic process with Markov property. The framework aims to capture the complexity of two-phase flow using the advantage of reinforcement learning (RL) in discovering complex patterns with the help of the physical model (Interfacial Area Transport Equation \cite{kocamustafaogullari1995foundation}) as reference. The details of the framework design are described including the design of the environment and the algorithm used in solving the RL problem \cite{lillicrap2015continuous}. The performance of the PIRLF is tested through experiments using the experimental database for vertical upward bubbly air-water flows \cite{wu1998one}. The result shows a good performance of PIRLF with rRMSE of 6.556\%. The case studies on the PIRLF performance also show that the type of reward function that is related to the physical model can affect the framework performance. Based on the study, the optimal reward function is established. The approaches to extending the capability of PIRLF are discussed, which can be a reference for the further development of this methodology.

\end{abstract}

\begin{keyword}


Reinforcement learning \sep Deep Deterministic Policy Gradient \sep Two phase flow \sep Interfacial area transport \sep Interfacial area concentration
\end{keyword}

\end{frontmatter}


\textbf{Nomenclature}
\vspace{3mm}

\textbf{$A$}  \quad    Action 

\textbf{$a_{i}$}  \quad  Interfacial area concentration (IAC)

\textbf{$D_{sm}$} \quad Bubble Sauter mean diameter

\textbf{$g$} \quad  Gravitational acceleration

\textbf{$P$} \quad Probability transition function

\textbf{$R$, $r$} \quad Rewards/reward function

\textbf{$R_{B}$} \quad Rate of change of IAC due to bubble coalescence%

\textbf{$R_{C}$}\quad  Rate of change of IAC due to bubble breakup%

\textbf{$R_{P}$} \quad Rate of change of IAC due to phase change%

\textbf{$S$, $s$} \quad  State/state set

\textbf{$u_{t}$} \quad Turbulent fluctuation velocity

\textbf{$v$}\quad  Bubble velocity

\textbf{$We$}\quad  Weber number

\vspace{3mm}
\textit{Greek letters}
\vspace{3mm}

\textbf{$\alpha$} \quad  Void fraction

\textbf{$\rho$}\quad Density

\textbf{$\sigma$}\quad Surface tension

\textbf{$\mu$}\quad Kinematic viscosity

\textbf{$\psi$}\quad Bubble shape factor

\vspace{3mm}
\textit{Subscripts}
\vspace{3mm}

\textbf{$exp$} \quad experiment

\textbf{$f$} \quad liquid

\textbf{$g$} \quad gas

\textbf{$IATE$} \quad prediction by IATE

\textbf{$pred$} \quad prediction by PIRLF

\textbf{$RC$} \quad Random collision

\textbf{$TI$} \quad Turbulent impact

\textbf{$WE$} \quad Wake entrainment

\vspace{3mm}
\textit{Abbreviations}
\vspace{3mm}

$DDPG$  \quad Deep Deterministic Policy Gradients

$IAC$ \quad Interfacial area concentration

$IATE$ \quad Interfacial area transport equation

$RL$ \quad Reinforcement Learning

$MDP$ \quad Markov Decision Process

$OU$ \quad Ornstein-Uhlenbeck function

$PIRLF$ \quad Physics-informed Reinforcement Learning framework

$PDE$ \quad Partial differential equation

$rRMSE$ \quad relative rooted-mean-square error

$ReLU$ \quad Rectified linear unit function 

$TRACE$ \quad TRAC/RELAP Advanced Computational Engine

\pagebreak

\section{Introduction}
The prediction of the characteristics of the two-phase flow is essential in terms of the safety of two-phase flow systems such as the reactor pressure vessel in the nuclear power plant. Nowadays, many software and codes have been developed based on two-phase flow fundamental theories and models. Take the TRACE code \cite{trace2007theory} as an example, which is considered as one of the most elaborated code up to now. The TRACE code was developed based on the Two-Fluid Model \cite{ishii1984two} and interfacial area transport equations (IATE) \cite{kocamustafaogullari1995foundation} and is capable of predicting the characteristics of the two-phase flows \cite{bernard2014implementation}. The Two-Fluid Model uses two groups of partial differential equations and a series of constitutive equations to describe the two phases. The IATE, which serves as the constitutive equations of interfacial area concentration for the Two-Fluid Model, is developed analogous to the Boltzmann transport equation and is capable of dynamically predicting the transition of the two-phase flows. Both of the two models are considered to be the most accurate model up to now. However, the models are complicated to solve since they include many non-linear PDEs. During the years, more correlations and models have been developed and further elaborate on the models, making the models even more complex to compute.

Another issue raises when the complex two-phase flow phenomena are modeled using deterministic correlations and models, which include approximations and averages. These deterministic models can hardly handle the two-phase flow with high turbulence and degree of freedom. One way to avoid this issue is to model the two-phase problem in the stochastic approach along with probability distribution functions. There are past studies in fluid mechanics that involve treatment of the deterministic basic equations of fluid mechanics via statistical and stochastic methods to solve the velocity field in turbulent flows \cite{durbin1980stochastic,novikov1989two,pedrizzetti1994markov}. They assume that the fluid flow has Markov property. It can be further thought that the fluid flow and two-phase flow with the Markov property can be modeled using Markov decision process and reinforcement learning problem.

A reinforcement learning problem is about learning from interaction how to behave to achieve a goal \cite{sutton1998introduction}. The reinforcement learning (RL) problem/setup constitutes a Markov decision process (MDP). An MDP is a discrete, stochastic control process that provides mathematical frameworks for decision making \cite{bellman1957markovian}. The key element of the MDP can be represented as a tuple, $(S, A, P, R)_{n}$. S, A, P, R are the state, action, probability transition function, rewards of the n-th agent, respectively. The agent interacts with the environment in terms of the state by taking actions and getting rewards. A policy, which is a stochastic rule by which the agent selects actions as a function of states, is formed through the process. The objective is to maximize the amount of reward it receives over time \cite{sutton1998introduction}. By projecting a two-phase flow problem into an MDP, which includes encoding the flow conditions and fluid properties into the state and associating the problem metric with the reward, the RL can provide solutions to the two-phase flow problem in a stochastic approach.

In this paper, a physics-informed reinforcement learning framework that predicts the interfacial area change with the aid of the IATE was proposed. This framework is based on the MDP that the changes and transitions of the two-phase flows meet the Markov property. In this paper, a detailed introduction of this framework and a comprehensive assessment of its performance are provided.


\section{Methodology}
\subsection{Physics-informed Reinforcement Learning Framework (PIRLF)}
Consider a steady-state two-phase flow in a finite length, vertical, round duct. There is no heat source existed in the duct and the flow is under atmospheric, adiabatic condition. Suppose that the well-mixed two-phase mixture entering into the duct from the bottom and passing through the duct is under a certain initial state: $\S = \{ \alpha_{0}, a_{i,0}, v_{g,0}, P_{0}, ... \}$. Due to the pressure, velocity change, and bubble interaction, the interfacial parameters of the two-phase flow change when the flow passes through the duct. The flow enters and exits the duct freely so that there is no entrance or exit effect. The goal is to predict the two-phase flow cross-sectional averaged interfacial area concentration when it passes through the duct. 

\begin{figure}[htbp] 
    \centering 
    \includegraphics[width=0.8\linewidth]{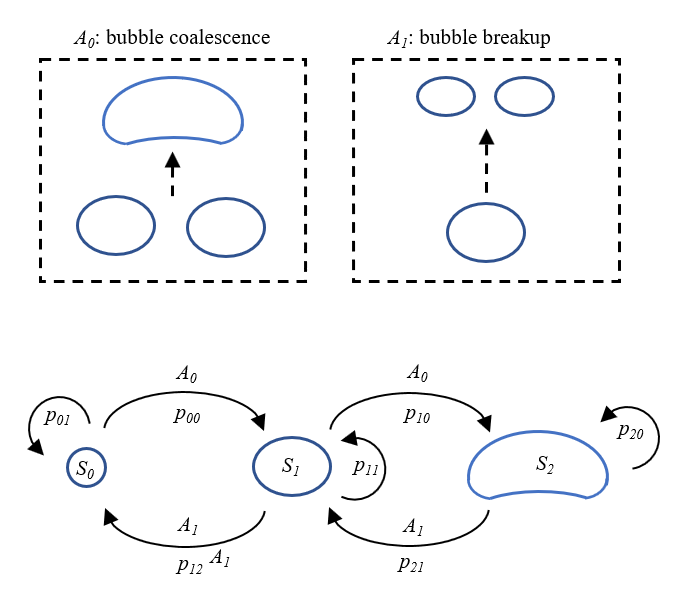} 
    \caption{Formulation of two-phase flow problem into a Markov Decision Process.} 
    \label{fig:MDP} 
\end{figure}

\begin{figure}[htbp] 
    \centering 
    \includegraphics[width=0.8\linewidth]{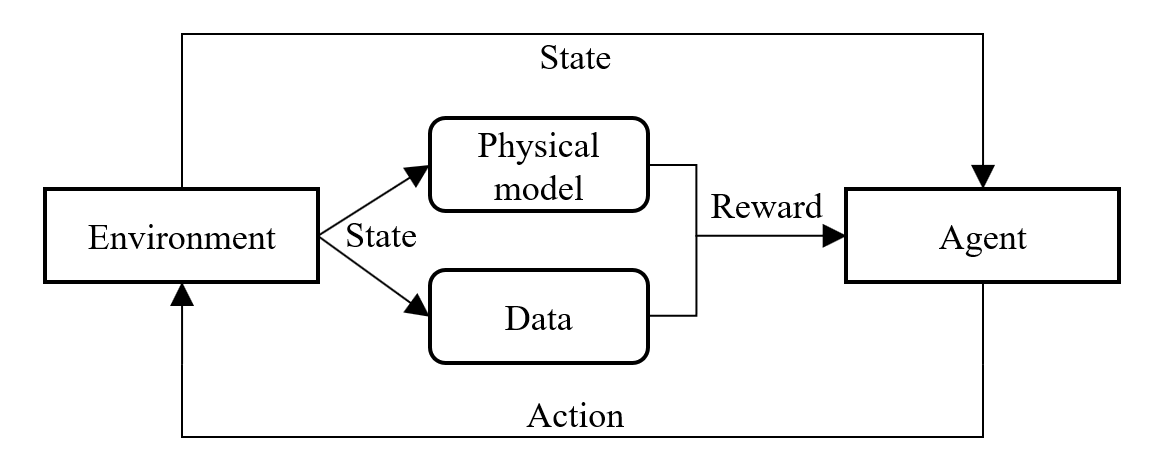} 
    \caption{The Physics-informed Reinforcement Learning.} 
    \label{fig:RL} 
\end{figure}

The above two-phase flow problem can be formulated into a Markov Decision Process (MDP). The motion of the bubbles in two-phase flow driven by the flow turbulence is partly stochastic, yet the overall bubble behaviors are partly determined. The two-phase flow problem projecting into an MDP is depicted in Fig. \ref{fig:MDP}. The figure simplifies the adiabatic, non-heated, air-water two-phase flow by considering only bubble interactions. Bubbles in the two-phase flow field could coalesce or break up depending on the states of the flow, such as bubble number density, bubble size, and flow turbulence, etc. The interaction next state of the bubbles (or the flow at a certain axial position) only depends on the current state, which satisfies Markov Property.
An MDP problem can be solved by reinforcement learning (RL). Given an MDP, the objective of the RL is to find an optimal policy by maximizing the rewards gained that gives actions based on the state. In this framework, the rewards are related to the accuracy of the interfacial area concentration prediction. Thus, an optimal policy means the framework can give good predictions on the interfacial parameters of the two-phase flow in the above problem. Fig. \ref{fig:RL} shows the conceptual diagram of the physics-informed RL framework. Specifically, the elements to establish an MDP including the state $S$, action $A$, transition probability function $P$, and the reward $R$ is described as follows.

\subsubsection{State, $S$}
The elements in the state are the basis of the agent to make decisions on which actions should be taken. In discrete state space that the elements of the state are discrete, the actions available at a certain state are usually also discrete. However, in this study, the action should be fine controlled to avoid discontinuities of the prediction. Therefore, the elements of the state and the action are both continuous, and the number of the state is infinite. The elements of the state consist of the hydrodynamic properties and the two-phase flow parameters. The principle of the state element selection is to cover all the influential factors that contribute to the prediction. Since some of the hydrodynamic properties are related and either can be used to describe a certain state, therefore, the selection of the elements of the state is not unique. Considering the current problem set aims on adiabatic, non-heated, air-water flow in a vertical tube, the air density $\rho_{g}$, void fraction $\alpha$, interfacial area concentration $a_{i}$, bubble velocity $v_{g}$, and bubble mean diameter $D_{sm}$ are selected to be the element of the state. Other two-phase flow properties that vary during the flow development are related to the above parameters.

\subsubsection{Action, $A$}
The action decides how to make a change on the state elements that can be changed. The possible action taken is determined based on the current state. In this problem setup, the bubble Sauter mean diameter is being changed by the agent based on the actions it chose. The main reason that bubble mean diameter is chosen as the tuning variable is that its upper and lower bounds can be approximately determined based on knowledge. The interfacial area concentration can be calculated based on the $D_{sm}$ and the void fraction $\alpha$,

\begin{equation}
a_{i}=\frac{6\alpha}{D_{sm}}
\end{equation}
The void fraction and gas density are changed corresponding to the change of the local pressure, which changes as the two-phase flow elevates in the duct. Given the fact that the local pressures at each elevation of the duct are known in the IATE calculations, the void fraction and gas density are calculated based on the local pressure provided based on the experimental data. The other elements in the state can be calculated by these elements that are changed by the agent and the general parameters in the game setup.


\subsubsection{Transition probability function, $P$}
The simplest way to set the transition probability function is to set it "deterministically",
\begin{equation}
    P_{a}\left(s, s^{\prime}\right) = P \left(s_{t+1}=s^{\prime} \mid s_{t}=s, a_{t}=a\right) = 1
\end{equation}
It means that when taken action $a_{t}=a$ at state $s_{t}=s$, the probability that the next state $s_{t+1}=s^{\prime}$ is 1 and there is no other type of next state. Another way to formulate the transition probability function is to consider updating the probabilities of the occurrence of each state given the current state and action, which is similar to Bayesian inference,
\begin{equation}
    P_{a}\left(s, s^{\prime}\right)= P \left(s_{t+1} = s^{\prime} \mid s_{t}=s, a_{t}=a\right)= \frac{p(a \mid s^{\prime}, s) p(s^{\prime} \mid s)}{p(a \mid s)}
\end{equation}
For example, in the framework design, given an action $a_{t}=a$ that the average bubble diameter increases, the probabilities of the next state that the bubble diameter increases, holds, or decreases can be calculated based on the prior probability (given state $s$, the probability that the next state is $s^{\prime}$) and the evidence that action $a$ is taken given the state $s$.

\subsubsection{Rewards, $R$}
In an RL framework, a reward should be given after the agent makes action and changes the state. In this framework, the interfacial area concentration (IAC) $a_{i}$ is related to the reward. Therefore, the design of the reward function is the key to the success of the establishment of the framework. In the current study, the reward function can be given in the following listed methods:

\textit{Type-1: Physics-informed reward function}
    
The interfacial area transport can be estimated based on the IATE. In the IATE, the change of IAC $\Delta a_{i,IATE}$ is calculated using the bubble interaction source terms. The detailed information of IATE and the models of bubble interaction source terms are provided in the following section, Physical models. The reward function can be expressed as,

\begin{equation}\label{eq:r1}
    r = 
    \left\{\begin{matrix}
    - |\Delta a_{i,pred}-\Delta a_{i,IATE}|& random(0,1) < \epsilon\\
    0 & otherwise 
    \end{matrix}\right.
\end{equation}
A tuning factor $\epsilon$ is added to regulate this rewards function on to what degree it gives feedback to the agent, namely the inform frequency. It can be understand as the probability that the reward function will give feedback to the agent on the action it picked.

\textit{Type-2: Data-aided reward function}

The predict IACs can be compared with the experimental data. In the experimental studies on the interfacial area transport \cite{dang2017experimental, dang2020two}, the interfacial parameters are measured at different axial locations. The measurement values can be compared with the predicting values at the same axial positions,

\begin{equation}\label{eq:r2}
    r = 
    \left\{\begin{matrix}
    - \mid a_{i,pred}- a_{i,exp} \mid & z_{pred}=z_{exp}\\
    0 & otherwise \\
    \end{matrix}\right.
\end{equation}

\textit{Type-3: Hybrid reward function}

This method can be considered as a combination of the last 2 methods.

\subsection{Physical models}
 
The axial change of the IAC is modeled using one-group interfacial area transport equation, which is expressed as \cite{wu1998one,hibiki2002development}
\begin{equation}\label{eq:iate}
    \begin{aligned}
\frac{\partial a_{\mathrm{i}}}{\partial t}+\frac{d}{\mathrm{d} z}\left(a_{\mathrm{i}} v_{\mathrm{g}}\right)=& \frac{1}{3 \psi}\left(\frac{\alpha}{a_{\mathrm{i}}}\right)^{2}\left(R_{\mathrm{B}}-R_{\mathrm{C}}+R_{\mathrm{P}}\right) \\
&+\left(\frac{2 a_{\mathrm{i}}}{3 \alpha}\right)\left\{\frac{\partial \alpha}{\partial t}+\frac{d}{\mathrm{d} z}\left(\alpha v_{\mathrm{g}}\right)\right\}
\end{aligned}
\end{equation}
where $\psi$ is the bubbles shape factor and equal to $1/36\pi$ for spherical bubbles. $R_{\mathrm{B}}$, $R_{\mathrm{C}}$, $R_{\mathrm{P}}$ are the rate of change of IAC due to bubble breakup, rate of change of IAC due to bubble coalescence, and rate of change of IAC due to phase change, respectively. Eq. \ref{eq:iate} can be simplified under steady-state adiabatic two-phase flow and converts into one-dimensional form by applying cross-sectional area averaging,
\begin{equation}\label{eq:1diate}
\frac{d}{\mathrm{d} z}\left(\left\langle a_{\mathrm{i}}\right\rangle \left\langle\left\langle v_{\mathrm{gz}} \right\rangle\right\rangle_{a} \right)=\left\langle\varPhi_{\mathrm{B}}\right\rangle-\left\langle\varPhi_{\mathrm{C}}\right\rangle+\frac{2\left\langle a_{\mathrm{i}} \right\rangle }{3\left\langle \alpha \right\rangle} \frac{\mathrm{d}}{\mathrm{d} z}\left(\left\langle \alpha  \right\rangle \left\langle\left\langle v_{\mathrm{g}} \right\rangle\right\rangle\right)
\end{equation}
where $\left\langle \; \right\rangle$, $\left\langle\left\langle \; \right\rangle\right\rangle_{a}$, and $\left\langle\left\langle \; \right\rangle\right\rangle$ denote the area-averaged quantity, interfacial area concentration-weighted mean quantity, and void fraction-weighted mean quantity, respectively. $\varPhi_{\mathrm{B}}$ and $\varPhi_{\mathrm{C}}$ represent the interfacial area concentration source terms due to bubble breakup and coalescence, respectively.  In the current one-dimensional formulation for adiabatic two-phase flow, the covariance terms are neglected because of relatively uniform flow parameter distribution over the flow channel \cite{hibiki2009interfacial}.
Wu et al. \cite{wu1998one} further simplified Eq. \ref{eq:1diate} by eliminating the third term on the right-hand side, assuming that the gas phase is incompressible without phase change. In bubbly flow, the bubble breakup and coalescence source terms were categorized by Wu et al. \cite{wu1998one} into three types: bubble coalescence due to random collision, bubble coalescence due to wake entrainment, and bubble breakup due to turbulent impact. The corresponding models are expressed as \cite{wu1998one},

\begin{equation}\label{eq:RC}
    \begin{aligned}
\varPhi_{\mathrm{RC}}=-\frac{1}{3 \pi} C_{\mathrm{RC}}(&\left.u_{\mathrm{t}} a_{i}^{2}\right)\left[\frac{1}{\alpha_{\max }^{1 / 3}\left(\alpha_{\max }^{1 / 3}-\alpha^{1 / 3}\right)}\right] \\
& \times\left[1-\exp \left(-C \frac{\alpha_{\max }^{1 / 3} \alpha^{1 / 3}}{\alpha_{\max }^{1 / 3}-\alpha^{1 / 3}}\right)\right]
\end{aligned}
\end{equation}
\begin{equation}\label{eq:WE}
    \varPhi_{\mathrm{WE}}=-\frac{1}{3 \pi} C_{\mathrm{WE}} u_{\mathrm{r}} a_{\mathrm{i}}^{2}
\end{equation}
\begin{equation}\label{eq:TI}
    \begin{aligned}
\varPhi_{\mathrm{TI}}=\frac{1}{18} C_{\mathrm{TI}} u_{\mathrm{t}} &\left(\frac{a_{i}^{2}}{\alpha}\right)\left(1-\frac{W e_{\mathrm{cr}}}{W e}\right)^{1 / 2} \\
& \times \exp \left(-\frac{W e_{\mathrm{cr}}}{W e}\right), \quad W e>W e_{\mathrm{cr}}
\end{aligned}
\end{equation}
where parameters such as $a_{i}$ and $\alpha$ are all averaged quantities in Eq. \ref{eq:RC}-\ref{eq:TI}, $u_{\mathrm{r}}$, $u_{\mathrm{t}}$, and $W e$ are calculated based on the averaged values as well. $C_{\mathrm{RC}}$, $C_{\mathrm{WE}}$, $C_{\mathrm{TI}}$ are coefficients that generated through assumptions and simplifications in the model development, which can be estimated based on experimental data. After applying the source terms, Eq. \ref{eq:1diate} becomes,
\begin{equation}\label{eq:finaliate}
    \frac{d}{\mathrm{d} z}\left(\left\langle a_{\mathrm{i}}\right\rangle \left\langle\left\langle v_{\mathrm{gz}} \right\rangle\right\rangle_{a} \right)=\varPhi_{\mathrm{RC}}+\varPhi_{\mathrm{WE}}+\varPhi_{\mathrm{TI}}
\end{equation}
Eq. \ref{eq:finaliate} calculates the change of IAC at each axial position, which corresponds to $\Delta a_{i,IATE}$ in Eq. \ref{eq:r1}.


\subsection{Reinforcement learning algorithm}

In this study, Deep Deterministic Policy Gradients (DDPG) \cite{lillicrap2015continuous} that features with continuous control is used to solve the reinforcement learning problem. The DDPG was developed based on 3 algorithms: Deterministic Policy-Gradient Algorithms \cite{silver2014deterministic}, Actor-Critic Methods \cite{bhatnagar2008incremental}, and Deep Q Network (DQN) \cite{mnih2015human}.
Similar to the Actor-critic Methods, an actor function is the policy function that models the expected action taken given the current state. The critic function is a Q-value function that the Q-values are updated based on the Bellman equation. In DDPG, both the actor and critic function are modeled by neural networks. The actor function gives deterministic action instead of a probability distribution on the action space. The critic and actor networks are updated in the off-policy approach that trains the neural network on batch experiences $(S, A, P, R)$ randomly picked in a replay buffer. Besides, as in DQN, the target networks corresponding to the actor and critic networks are also used in this algorithm to store the past networks. The purpose of using the target network is to stabilize the learning process.

\subsection{Summary of the framework}
\begin{figure}[htbp] 
    \centering 
    \includegraphics[width=0.8\linewidth]{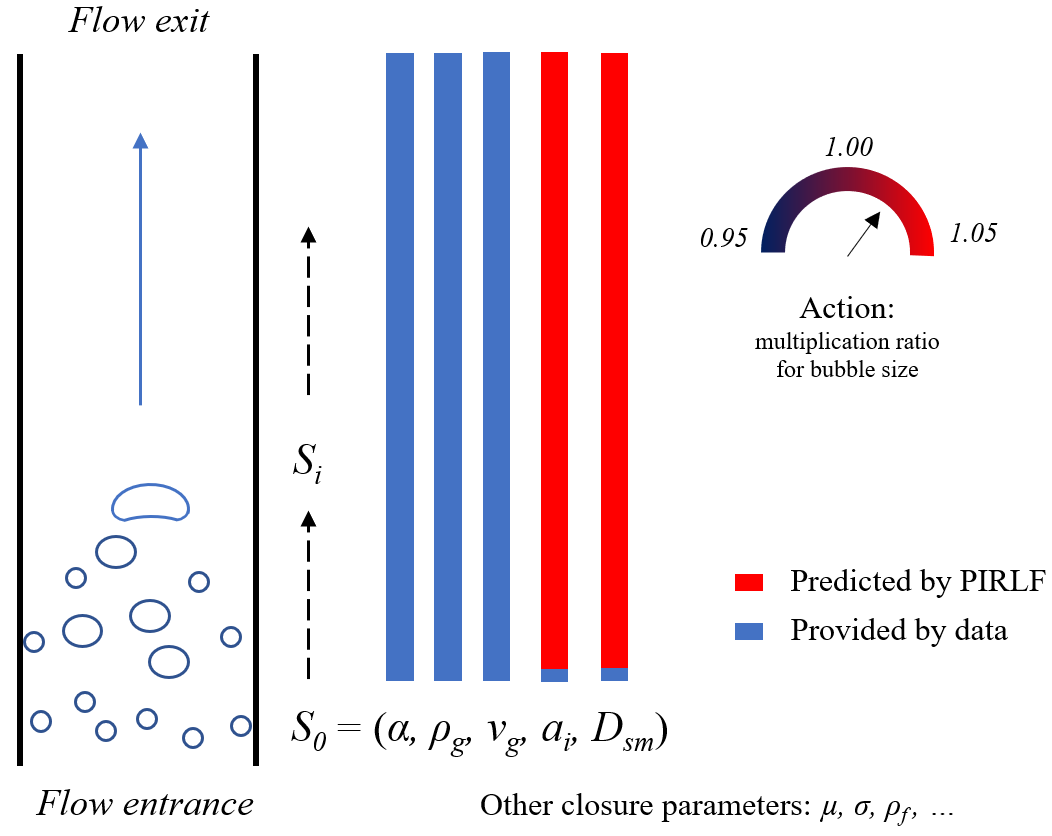} 
    \caption{The MDP environment that describes the interfacial area change for vertical upward air-water two-phase flow.} 
    \label{fig:env} 
\end{figure}

The Physics-informed Reinforcement Learning Framework (PIRLF) for the interfacial area transport in two-phase flow consists of two parts: one is to project the two-phase flow problem into an MDP; the other one is to use the advantage of reinforcement learning to solve the MDP problem. The first part can be considered as creating an environment that is utilized by a reinforcement learning algorithm for exploration and exploitation, which is included in the second part. Fig. \ref{fig:env} summarizes the environment design for this problem. The ultimate goal is to find the optimal policy that gives satisfying estimations on the interfacial area transport. Specifically, the rules of state transformation and the selected action, which include the information of the two-phase flow, should be properly included in the environment. The reward function should be related to the purpose of establishing the framework. The selection of the reinforcement learning algorithm should be able to solve this problem that has continuous state and action space.

\section{Experiment}
This study primarily focuses on the performance of the framework in predicting the interfacial area transport in bubbly flow. Since this study uses the model developed by Wu et al. \cite{wu1998one}, 7 flow conditions proposed in Wu et al.'s work are utilized as the database to train and test the framework.
Specifically, one of the flow conditions is randomly selected as the test case while the remaining of the flow conditions are used as the training test cases. This is similar to the k-fold cross-validation method (k=7 in this case). The objective of the training process is to find the optimal policy using the DDPG algorithm that gives satisfying estimations on the IAC changes through exploration and exploitation, based on the experimental data included in the training set.

\begin{figure}[htbp] 
    \centering 
    \includegraphics[width=0.7\linewidth]{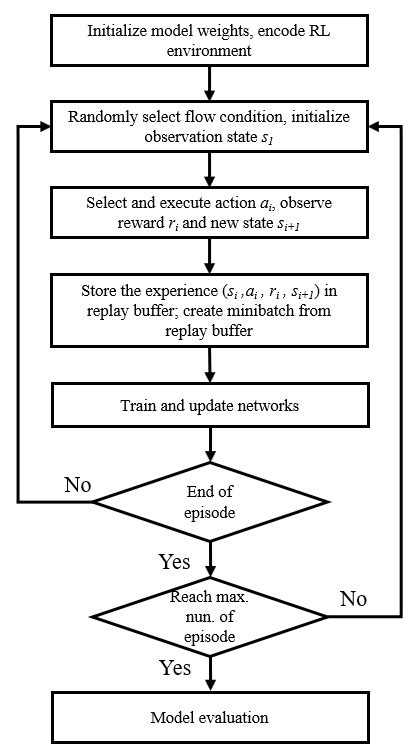} 
    \caption{The implementation flow of the framework.} 
    \label{fig:algorithm} 
\end{figure}

The detailed procedure of model training is provided in Fig. \ref{fig:algorithm}
Resemble the DDPG algorithm \cite{lillicrap2015continuous}, the training procedure includes model initialization, exploration of the environment in episodes, and updating of the networks. Before the training of the DDPG model, the networks are initialized randomly to avoid the possibility of divergence. During the training process, each episode consists of a fixed number of steps that correspond to the number of discretized axial positions along the vertical tube. One of the 6 training flow conditions is randomly selected. The initial state that contains initial values of IAC, void fraction, etc. are provided based on the experimental measurements of the selected flow condition. For each step in one episode, the action is to change the bubble Sauter mean diameter by providing a ratio in a continuous domain. In the DDPG, the action picked in a step is based on the output of the actor network plus some exploration noise. The bubble Sauter mean diameter as well as IAC will be changed based on the ratio selected by the action. The other parameters, such as void fraction and bubble velocity, are changed based on the boundary conditions provided by the experimental measurements. The experience is saved in a replay buffer for training and updating the networks. The actor and critic networks as well as the target networks are trained and updated using the minibatch data, randomly selected in the replay buffer. The detailed methods on how to train and update the networks can be found in the original paper \cite{lillicrap2015continuous}). In the model evaluation part, the performance will be tested using the test data set. The performance of the framework is quantified using the relative rooted-mean-square error (rRMSE),
\begin{equation}
    r R M S E=\sqrt{\frac{1}{n} \sum_{i=1}^{n}\left(\frac{a_{i,pred}^{(i)}-a_{i,exp}^{(i)}}{a_{i,exp}^{(i)}}\right)^{2}}
\end{equation}

\section{Analysis of the performance}

\subsection{Effect of reward function type}

The axial change of IAC is mainly due to three effects: the pressure effect that causes gas expanding along the tube, the advection effect caused by the bubble velocity change, and bubble interaction that includes bubble coalescence and breakup. The first two effects cause steady IAC change since there is less likely to have a sudden change of pressure or bubble velocity under the current bubbly flow conditions and flow geometry. The bubble interaction could have some randomness since the bubble interactions are driven by the flow turbulence. However, statistically, the bubble interactions should follow some certain trends, thus, the axial change of IAC flow should be steady. Therefore, although the axial change of IAC along the tube can be either increase or decrease, or both could be observed, the trend of the change should be steady.

\begin{figure}[htbp]
    \centering 
    \includegraphics[width=1.0\linewidth]{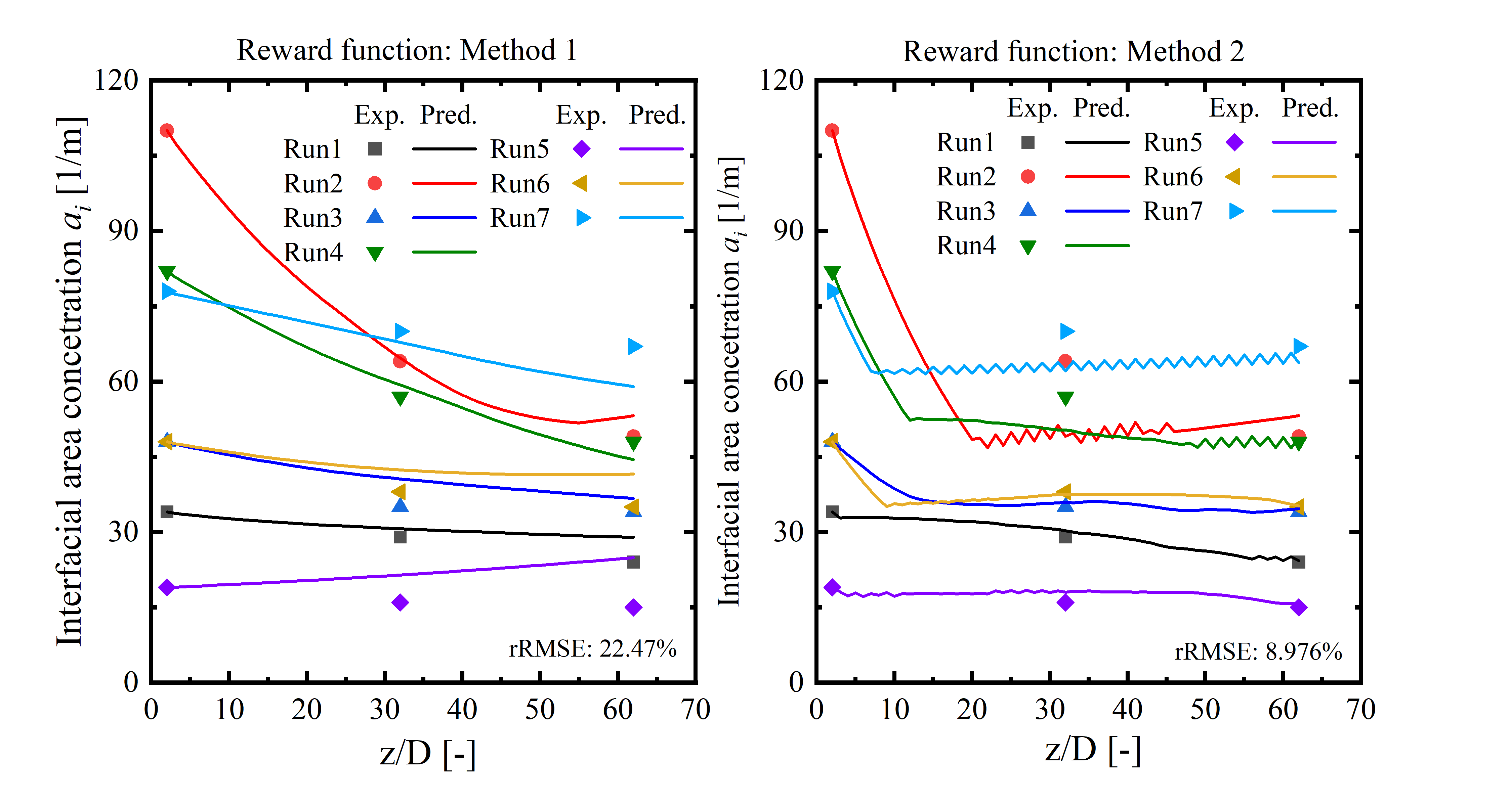} 
    \caption{Prediction results against experimental data by PIRLF with left): physics-informed reward function; right): data-aided reward function.} 
    \label{fig:effectrf} 
\end{figure}

Given the above discussion on the axial change of IAC, it can be concluded that besides the accuracy of IAC predictions, the steadiness of the predicting IAC curve is also an important criterion to analyze the performance of the model. Fig. \ref{fig:effectrf} shows the PIRLF prediction results with two standalone reward functions, physics-informed reward function (Type-1) and data-aided reward function (Type-2).
The framework with the Type-1 reward function is not as good as that with the Type-2 reward function on giving accurate predictions compared with experimental measurements. However, the predicting curve is more steady and reasonable. This is because the physics-informed reward function uses the mechanistic models which contain uncertainties, while the experimental data that is used in the data-aided reward function are regarded as true values. However, the mechanistic models can give steady IAC predictions, which serve as references to the framework and can result in steady predictions of the framework.
On the other hand, the data-aided reward function can lead to good predictions of IAC comparing with the data, however, the predictions can oscillate within the regions between the measurement positions. This is due to that the reward function doesn't provide information on these regions. The optimal policy can choose any actions within the action space as long as it can give good predictions at the position where the measurements are taken. Therefore, the change of IAC predicted may not be as steady as the IATE predicts between the measurement positions.

\begin{figure}[htbp]
    \centering 
    \includegraphics[width=0.7\linewidth]{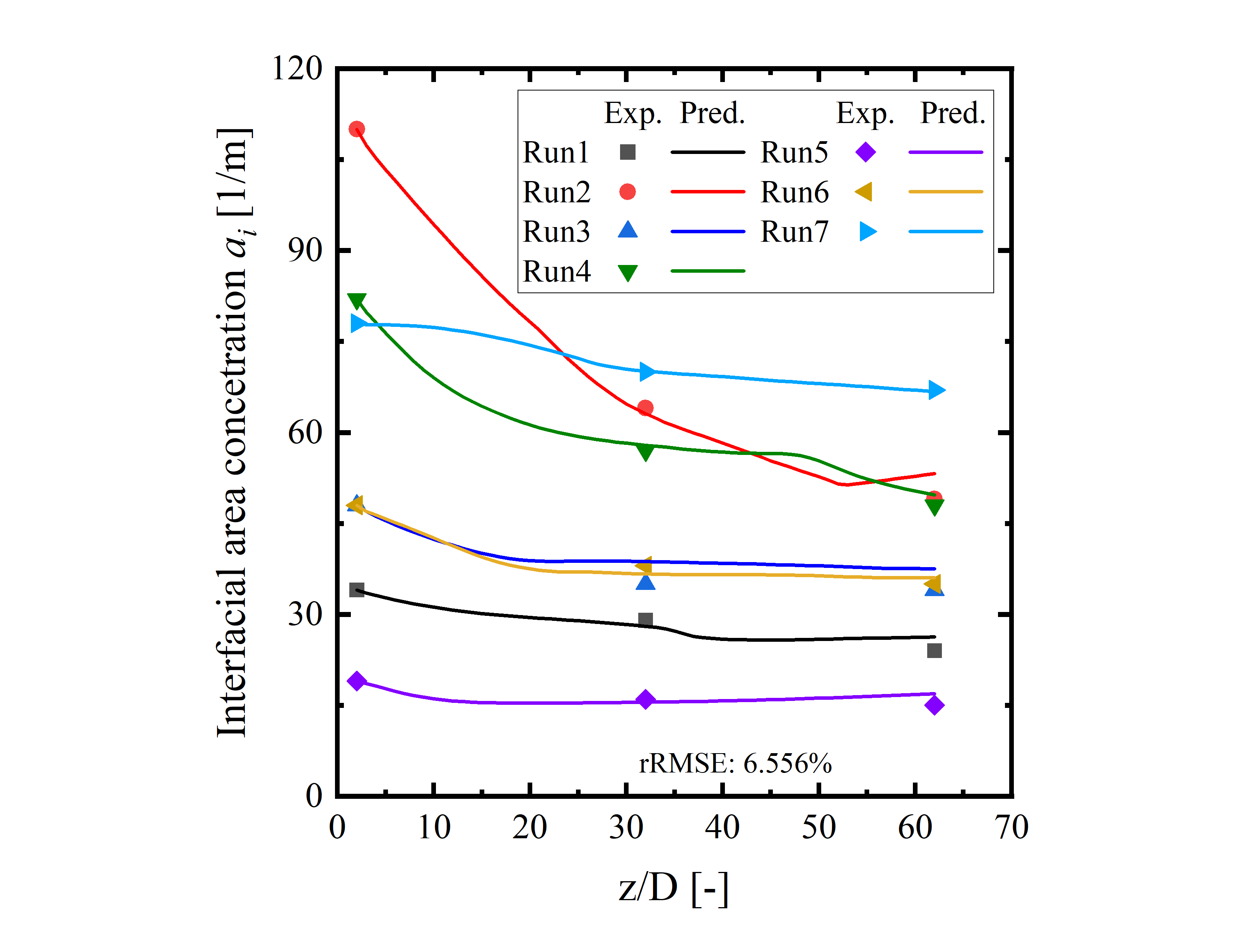} 
    \caption{Prediction results against experimental data by PIRLF hybrid reward function.} 
    \label{fig:hybridrf} 
\end{figure}

To take the advantages of the two types of reward functions to realize accurate and steady predictions, the hybrid reward function (Type-3) that includes both physical models and experimental data is applied. The performance of the framework with this hybrid reward function is given in Fig. \ref{fig:hybridrf}. Comparing the results with those given in Fig. \ref{fig:effectrf}, the PIRLF with hybrid reward function not only gives improved predictions compared with the results by data-aided reward function but also keeps the steadiness of prediction curves like physics-informed reward function.

\subsection{Sensitivity study on the inform frequency, $\epsilon$}

\begin{figure}[htbp]
    \centering 
    \includegraphics[width=0.7\linewidth]{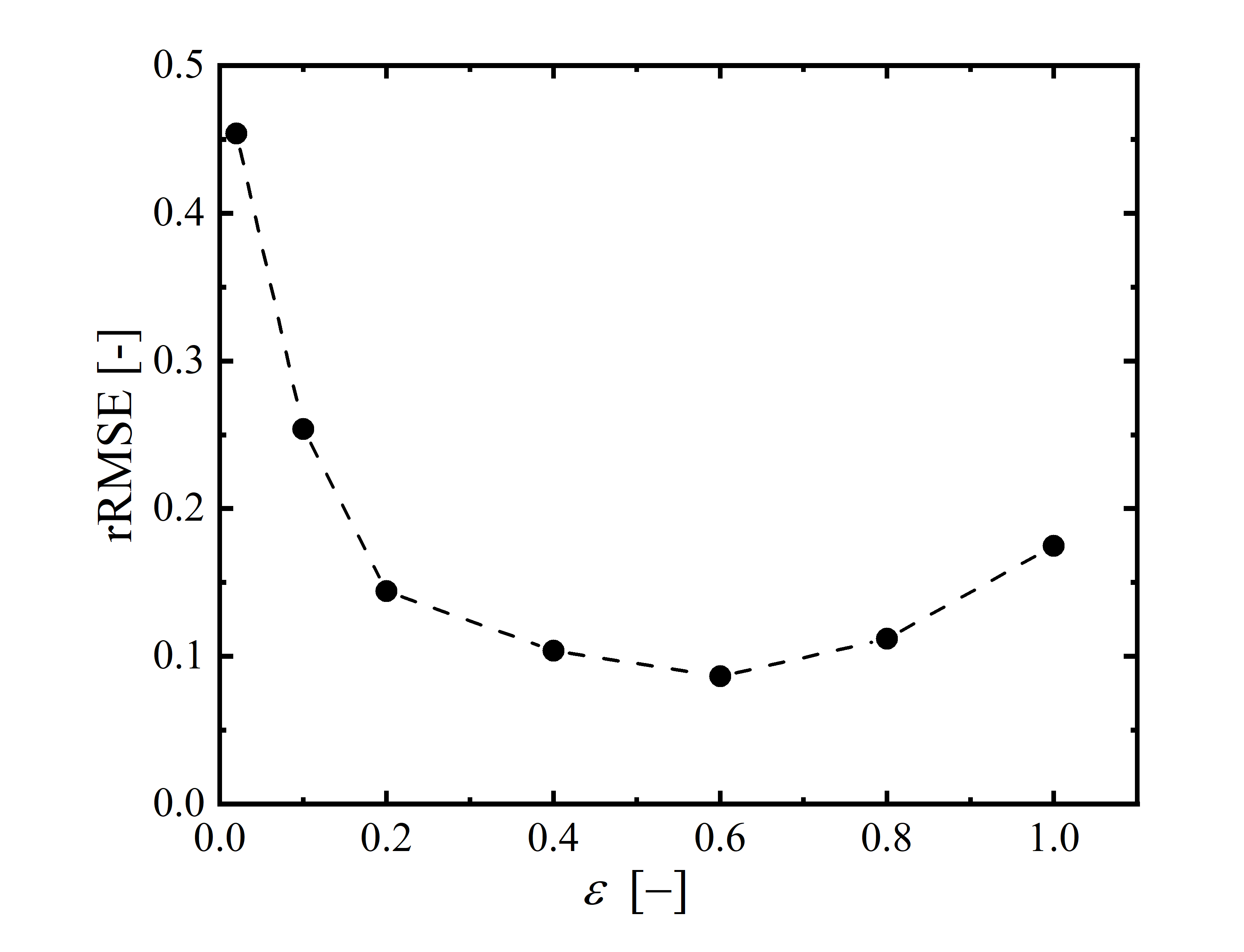} 
    \caption{Prediction results against experimental data by PIRLF hybrid reward function.} 
    \label{fig:eps} 
\end{figure}

In the physics-informed or hybrid reward function, the inform frequency, $\epsilon$, defined in Eq. \ref{eq:r1} is a key parameter that affects the PIRLF performance. Fig. \ref{fig:eps} shows the framework performance presented by rRMSE against $\epsilon$. The rRMSEs are generated using the PIRLF with physics-informed reward function (Type-1 reward). From the figure, there is an optimal $\epsilon$ that minimizes the framework prediction error. A large $\epsilon$ means the agent in PIRLF can frequently receive feedback from the reward function on how close its prediction to the IATE models. The PIRLF predictions are close to the IATE predictions, meaning they almost have the same prediction errors.  With a small $\epsilon$, the agent can hardly get feedback about its performance. Its selected actions can be quite random without capturing the characteristics contained in the environment. The results in the figure show that when $\epsilon$ is around 0.6, the PIRLF gives the best prediction results. In this situation, the agent can receive enough feedback from the reward function, and it also has some degree of freedom for exploration. Combined with the previous section 4.1, it can be concluded that the optimal reward function should be the hybrid-type reward function with the inform frequency $\epsilon$ of around 0.6.

\section{Towards extending the PIRLF capability}
The capability of the PIRLF in giving predictions on interfacial area transport depends on several aspects. In this section, the potential approaches to extending the PIRLF capability on predicting interfacial area transport are discussed, which could be references to further studies.

Firstly, as discussed in the previous section, the performance of the PIRLF depends on the data and physical models. In the analysis of the PIRLF performance in this study, the database only covers adiabatic, air-water bubbly flow in 50.8mm ID vertical tube. Besides, the IATE by Wu et al. \cite{wu1998one} was developed specifically for bubbly flow. By adding data with a broader range of flow conditions and flow geometries \cite{hibiki2001axial,ozar2013interfacial,dang2017experimental,du2020experimental,dang2020two}, the environment of the framework could include more characteristics of the interfacial area transport. Thus, the framework could be more comprehensive by finding optimal policies that can deal with a wider range of flow conditions and flow geometries. The selection of the IATE model, which serves as part of the reward function and provides references for finding the optimal policies, can also be improved by using more complex versions with broad predicting capabilities \cite{fu2003two, sun2001two, worosz2015interfacial}.

Besides, the environment setup is also considered to influence PIRLF performance and capability.  Besides the form of the reward function, the way to encode the state set can also be a non-negligible factor. This is because the state set is the input of the networks of the RL algorithm that aims to extract the patterns from the features included in the state set. Firstly, the information included in the state set should be enough to fully describe the characteristics of the interfacial area transport. For example, in order to give satisfying predictions on subcooled boiling two-phase flow, additional parameters such as imposed heat flux, fluid temperatures, etc. are also necessary to include in the state set. Secondly, since the orders of magnitude of the parameters in the state set are different (e.g. $a_{i}: 10^{2}$; $\alpha: 10^{-1}$), and neural networks are used in the RL algorithm, it is necessary to perform proper normalization on the state set to ensure good results.

Lastly, tuning the hyperparameters such as the numbers of network layers and perceptrons in the RL model can also help improve the predicting accuracy, but not substantially. Other RL algorithms that meet the problem requirements could also be used such as Soft Actor-Critic (SAC) \cite{haarnoja2018soft}.


\section{Conclusion}

This study proposed a novel physics-informed reinforcement learning-aided framework (PIRLF) aiming at predicting the interfacial area transport in two-phase flow. The two-phase flow problem is projected into a Markov Decision Process as the bubble transport that is mainly driven by turbulence is considered to be partly stochastic. The important information on designing the environment for Reinforcement Learning (RL) are summarized as follows:
\begin{itemize}
    \item The state of the environment is the basis to make decisions on how the bubble transport happens. The state set includes the hydrodynamic and two-phase flow parameters that are related to bubble transport. Meanwhile, parameters that are not related to the transport should not be included in the state set. Given the flow condition focused in this study, the air density $\rho_{g}$, void fraction $\alpha$, interfacial area concentration $a_{i}$, bubble velocity $v_{g}$, and bubble mean diameter $D_{sm}$ are selected to be the element of the state.
    \item The action decides how to make a change on the state elements that can be changed. The possible action taken is determined based on the current state. The action in the PIRLF is to change the bubble Sauter mean diameter by a multiplication ratio selected in a continuous space.
    \item The reward function should be related to the purpose of establishing the framework. In this study, 3 different types of reward functions are proposed: physics-informed, data-aided, and hybrid reward function. 
\end{itemize}

The Deep Deterministic Policy Gradients (DDPG) \cite{lillicrap2015continuous} algorithm is used to solve the two-phase flow RL problem. The performance of the PIRLF is tested through experiments using the experimental database for vertical upward bubbly air-water flows \cite{wu1998one}. The result shows by using a hybrid reward function, a satisfyingly good performance is achieved by PIRLF with rRMSE of 6.556\%. The PIRLF performance is largely related to the type of reward function as well as the performance of the physical model. This study also shows there is an optimal inform frequency $\epsilon$, which determines the possibility that the framework gets feedback from the reward function. The approaches to extending the capability of PIRLF are discussed, which can be a reference for the further development of this methodology.

\appendix

\section{Supplement information on the DDPG training}

As suggested by the original DDPG paper \cite{lillicrap2015continuous}, Adam optimizer \cite{kingma2014adam} is applied for learning the parameters in the actor and critic neural networks with learning rates of $10^{-4}$ and $10^{-3}$, respectively. For the value calculation using the action-value function, the discount factor $\gamma=0.99$. The activation function used in the hidden layers of actor and critic neural networks is the rectified linear unit function (ReLU), and the output layer of actor and critic neural networks are tanh and linear function. The actor network uses three hidden layers with 50 units each. The setup of other parameters follows the work by \cite{lillicrap2015continuous}. The exploration action noise is generated using Ornstein-Uhlenbeck function with $\theta=0$ and $\sigma=0.5$.



\bibliographystyle{elsarticle-num} 
\bibliography{references}

\end{document}